\documentclass[letter]{aa} 

\usepackage[english]{babel}
\usepackage{amssymb}
\usepackage{amsmath}
\usepackage{mathrsfs}
\usepackage{latexsym}
\usepackage{longtable}
\usepackage{multirow}
\usepackage{epsf}
\usepackage{color}
\usepackage{graphicx}
\usepackage{float}
\usepackage{enumerate}
\usepackage{txfonts}
\usepackage{natbib}
\usepackage{comment}
\usepackage{capt-of}
\usepackage{rotating}

\begin{document} 
   \title{Asymmetric features in the protoplanetary disk  MWC~758 \thanks{Based on observations performed with SPHERE/VLT under program ID 60-9389(A)}}

   \author{M.~Benisty\inst{1}  \and
A.~Juhasz\inst{2}
\and 
A.~Boccaletti\inst{3}
\and
H.~Avenhaus\inst{4}
\and
J.~Milli\inst{5}
\and
C.~Thalmann\inst{6}
\and 
C.~Dominik\inst{7}
\and
P.~Pinilla\inst{8}
\and
E.~Buenzli\inst{9}
\and
A.~Pohl\inst{9,10}
\and 
J.-L.~Beuzit\inst{1}
\and
T.~Birnstiel\inst{11}
\and
J.~de~Boer\inst{5,8}
\and
M.~Bonnefoy\inst{1}
\and
G.~Chauvin\inst{1}
\and
V.~Christiaens\inst{4}
\and
A.~Garufi\inst{6}
\and 
C.~Grady\inst{12}
\and
T.~Henning\inst{9}
\and
N.~Huelamo\inst{13}
\and
A.~Isella\inst{14}
\and
M.~Langlois\inst{15}
\and
F.~M\'enard\inst{16,1}
\and
D.~Mouillet\inst{1}
\and
J.~Olofsson\inst{9}
\and
E.~Pantin\inst{17}
\and
C.~Pinte\inst{16,1}
\and
L.~Pueyo\inst{18}}

   \institute{Universit\'{e} Grenoble Alpes, IPAG, F-38000 Grenoble, France;  CNRS, IPAG, F-38000 Grenoble, France;\\
              \email{Myriam.Benisty@obs.ujf-grenoble.fr}
              \and
              Institute of Astronomy, Madingley Road, Cambridge CB3 OHA, United Kingdom
              \and
              LESIA, Observatoire de Paris, CNRS, Universit\'e Pierre et Marie Curie Paris~6, Universit\'e Denis Diderot Paris~7, 5 place Jules Janssen,  92195 Meudon, France    
              \and
              Departamento de Astronom\`ia, Universidad de Chile, Casilla 36-D, Santiago, Chile  
              \and
              ESO, Alonso de C\'ordova 3107, Vitacura, Casilla 19001, Santiago de Chile, Chile
              \and
              Institute for Astronomy, ETH Zurich, Wolfgang-Pauli-Strasse 27, 8093 Zurich, Switzerland   
              \and
             Sterrenkundig Instituut ÒAnton PannekoekÓ, Science Park 904, 1098 XH Amsterdam, The Netherlands
              \and
              Leiden Observatory, Leiden University, P.O. Box 9513, 2300RA Leiden, The Netherlands
              \and
              Max Planck Institute for Astronomy, K\"onigstuhl 17, D-69117 Heidelberg, Germany 
              \and
              Institute of Theoretical Astrophysics, Heidelberg University, Albert-Ueberle-Strasse 2, D-69120 Heidelberg, Germany
              \and
              Harvard-Smithsonian Center for Astrophysics, 60 Garden Street, Cambridge, MA 02138, USA
              \and
              Eureka Scientific and Goddard Space Flight Center, Code 667, Goddard Space Flight Center, Greenbelt, MD 20771, USA 
              \and
              Centro de Astrobiolog\'ia (INTA-CSIC); ESAC Campus, P.O. Box 78, E-28691 Villanueva de la Canada, Spain
              \and
              Department of Physics \& Astronomy, Rice University, 6100 Main Street, Houston, TX, 77005, USA
              \and
              Observatoire de Lyon, Centre de Recherche Astrophysique de Lyon, Ecole Normale Sup\'erieure de Lyon, CNRS, Universit\'e Lyon~1, UMR 5574, 9 avenue Charles Andr\'e, Saint-Genis Laval, 69230, France
              \and
	   UMI-FCA, CNRS/INSU, France (UMI 3386), and Dept. de Astronom\'{\i}a, Universidad de Chile, Santiago, Chile
	   \and  
	   Laboratoire AIM, CEA/DSM - CNRS - Universit\'e Paris Diderot, IRFU/SAp, 91191 Gif sur Yvette, France
	   \and
	   Space Telescope Science Institute, 3700 San Martin Drive, Baltimore, MD 21218 USA}

   \date{Received 3 March 2015 / Accepted 18 May 2015}

  \abstract
   {The study of dynamical processes in protoplanetary disks is essential to understand planet formation. In this context, transition disks are prime targets because they are at an advanced stage of disk clearing and may harbor direct signatures of disk evolution. }
   {We aim to derive new  constraints on the structure of the transition disk MWC~758,  to detect non-axisymmetric features and understand their origin. }
   {We obtained infrared polarized intensity observations of the protoplanetary disk MWC~758 with SPHERE/VLT at 1.04~$\mu$m to resolve scattered light at a smaller inner working angle (0.093\arcsec{}) and a higher angular resolution (0.027\arcsec{}) than previously achieved.}
   {We observe polarized scattered light within 0.53\arcsec{} (148~au) down to the inner working angle (26~au) and detect distinct non-axisymmetric features but no fully depleted cavity. The two small-scale spiral features that were previously detected with HiCIAO are resolved more clearly, and new features are identified, including two that are located at previously inaccessible radii close to the star. We present a model based on the spiral density wave theory with two planetary companions in circular orbits. The best model requires a high disk aspect ratio ($H/r\sim$0.20 at the planet locations) to account for the large pitch angles which implies a very warm disk.}
   {Our observations reveal the complex morphology of the disk  MWC~758.  To understand the origin of the detected  features, the combination of high-resolution observations in the submillimeter with ALMA and detailed modeling is needed.}

   \keywords{              }

   \maketitle
%

\section{Introduction}
The planetary systems detected so far show a great diversity in their architectures and natures  which may originate in the variety of initial conditions of their formation \citep{mordasini2012}. It is thus of fundamental importance to study the physical conditions in the circumstellar disks in which they form, and the processes that rule their evolution. Direct-imaging observations at high angular resolution of young disks provide an efficient way to constrain these processes because they may leave clear signatures in the disk structure. As an example, planet-disk interactions will naturally generate asymmetric features, such as vortices, spiral arms or eccentric gaps, and may be used to indirectly detect planets embedded in their parent disk \citep[e.g][]{kley2012, baruteau2014}. In this context, disks with gaps or cavities, called transition disks, are prime targets to study because they may be at an advanced stage of disk evolution \citep{espaillat2014}. Asymmetric features have already been observed in transition disks with high-contrast observations in the infrared (IR), which probe the disk surface layers \citep[e.g.,][]{hashimoto2011,muto2012, rameau2012, casassus2012, boccaletti2013, canovas2013, garufi2013, avenhaus2014,thalmann2014}.  Some of these features were traced deeper in the disk with CO submillimeter imaging \citep[e.g.,][]{christiaens2014}. 

In this context, the transition disk around the Herbig A5 star MWC~758 (HD~36112) is particularly interesting.  It is 3.5$\pm$2~Myr old \citep{meeus2012}, and is located near the edge of the Taurus star-forming region. The revised Hipparcos parallax data provide a distance of 279$^{+94}_{-58}$~pc \citep{vanleeuwen2007}.  
It is most likely a single star, as any stellar companion down to a mass limit of 80~$M_{\rm{Jup}}$ was ruled out based on sparse aperture masking observations \citep{grady2013}.  Submillimeter observations show a circumstellar disk extending up to 385$\pm$26 astronomical units (au) \citep{chapillon2008}, with an inclination and position angle of 21$^\circ$$\pm$2$^\circ$ and 65$^\circ$$\pm$7$^\circ$, respectively, as determined from the CO line \citep{isella2010}. Modeling indicates that the surface density steeply increases  between 40 and 100\,au, which supports the idea that the disk might be gravitationally perturbed by one or more companions.  The observations marginally resolved a cavity with a radial extent in the submillimeter estimated between 70 and 100\,au \citep{isella2010,andrews2011}. 
Using $Ks$ (2.15\,$\mu$m) direct imaging and $H$-band (1.65\,$\mu$m) polarimetric imaging with the High Contrast Instrument for the Subaru Next Generation Adaptive Optics (HiCIAO), \citet{grady2013} found no evidence for a cavity in scattered light down to  0.1\arcsec{} (28\,au) from the star and detected two small-scale spiral arms. The authors suggested that the southeast spiral might be launched by a 5$^{+3}_{-4}\,M_{\rm{Jup}}$  companion at 1.55\arcsec{} (432\,au; beyond the submillimeter dust emission) in a dynamically warm disk. 

In this Letter, we report IR polarized intensity observations of the transition disk MWC~758 obtained with the SPHERE instrument in the Y~band  (1.04\,$\mu$m).
The use of an extreme adaptive optics system, combined with high-contrast  coronagraphic and differential imaging techniques led to deeper observations, at a smaller inner working angle (IWA; 0.093\arcsec{}, 26\,au) and a higher angular resolution (0.027\arcsec{}, 7.5\,au) than previously achieved. 

\begin{figure*}[t]
   \centering
   		\includegraphics[width=1.05\textwidth]{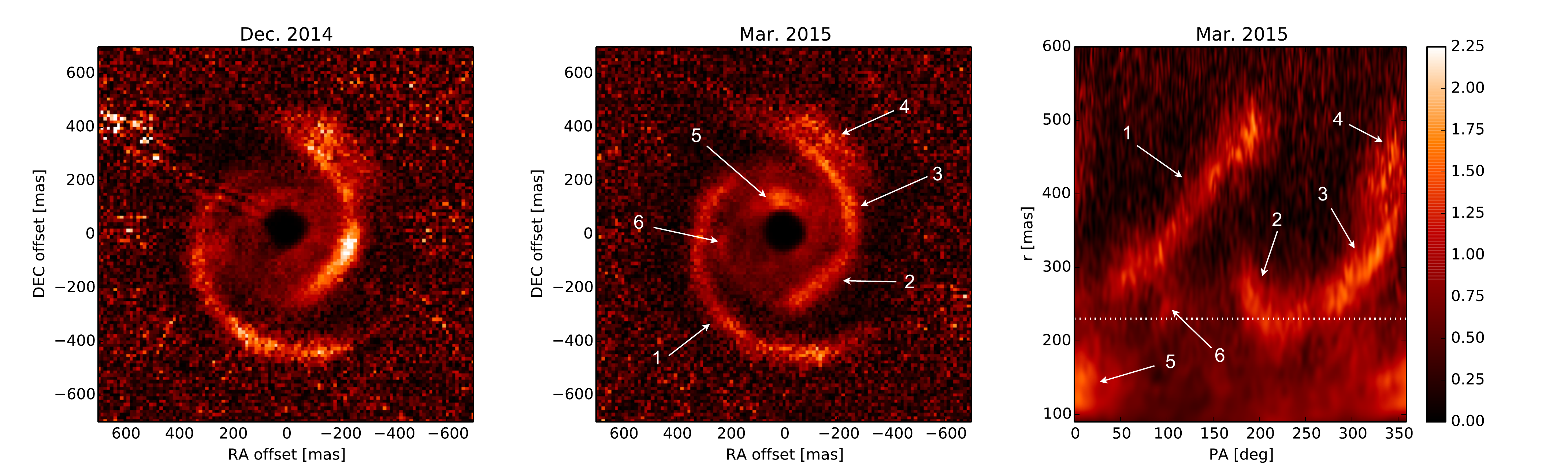} 
	\caption{Left and middle: polarized intensity images ($Q_{\phi}$) obtained in December 2014 and March 2015, respectively. East is toward the left. Right: radial map of the deprojected  $Q_{\phi}$ image from March 2015 using $i=21^\circ$ and $PA=65^\circ$. The dashed line indicates a radius of 0.23\arcsec{}. Each pixel has been scaled with the square of its distance from the star, $r^{2}$, to compensate for the $r^{-2}$ dependency of the stellar illumination.  The color scale is arbitrary. } 
	\label{fig:Qphi}
\end{figure*}

\section{Observations and data reduction}
The observations were carried out on 2014 December 5 and 2015 March 3 with the SPHERE instrument \citep[][Beuzit et al. in prep.]{sphere2008}, equipped with an extreme adaptive-optics (AO) system \citep{fusco2006,petit2014,sauvage2014}  at the Very Large Telescope at Cerro Paranal, Chile. The observations were executed in Science Verification Time. MWC~758 was observed in the $Y$-band filter (1.04~$\mu$m) using the infrared dual-band imager and spectrograph  \citep[IRDIS;][]{dohlen2008, maud2014}, with a plate scale of 12.25~milli-arcseconds (mas) per pixel. We used a 185~mas-diameter coronagraphic mask combined with a Lyot apodizer. 
MWC~758 was observed during $\sim$51 minutes at both epochs. These data were taken in challenging AO  conditions with moderate to poor seeing (between 0.9\arcsec{} and 1.2\arcsec{} in the optical), high airmass (between 1.5 and 2), and relatively short wavelength. 
The Strehl ratio reached about $\sim$70\% in the $H$~band, that is, $\sim$45\% in the $Y$ band. The December dataset suffered from a number of defects (regarding spider alignment and centering onto the mask), leading to a lower signal-to-noise   ratio (SNR) than in the March dataset.  

The images were obtained using differential polarimetric imaging (DPI), a powerful technique for studying the spatial distribution of small dust grains in the disk surface layers through their scattered light by suppressing the unpolarized stellar light \citep[e.g.][]{apai2004,quanz2011}. 
IRDIS provides two beams, in which wire-grid polarizers are inserted in DPI mode, and lead to ordinary and extraordinary polarization states. 
The half-wave plate (HWP) that controls the orientation of the polarization was set to four positions shifted by 22.5$^\circ$ in order to construct a set of linear Stokes vectors. To minimize uncertainties related to the relative positioning of the star with respect to the coronagraphic mask, the frame registration was accurately performed using a background star located at $\sim$2.35\arcsec{} from MWC~758. 
The data were corrected for the distortion and true North. We then followed a procedure developed for high-contrast imaging and used the double-ratio method to calculate the Stokes parameters $Q$ and $U$ \citep{avenhaus2014}.  Since the scattered light from a circumstellar disk is expected to be linearly polarized in the azimuthal direction under the assumption of single scattering, it is beneficial to describe the polarization vector field in polar rather than Cartesian coordinates.  We define the
polar-coordinate Stokes parameters $Q_\phi$, $U_\phi$ as:

\begin{equation}
Q_\phi = +Q \cos(2\Phi) + U \sin(2\Phi)
\end{equation}
\begin{equation}
U_\phi = -Q \sin(2\Phi) + U \cos(2\Phi), 
\end{equation}

\noindent where $\phi$ refers to the azimuth in polar coordinate, $\Phi$ is the position angle of the location of interest (x, y) with respect to the  star location (x$_{0}$,y$_{0}$), and is written as:
\begin{equation}
\Phi = \arctan\frac{x - x_{0}}{y-y_{0}} + \theta.
\end{equation}

$\theta$ corrects for instrumental effects such as the angular misalignment of the HWP. In this coordinate system, the azimuthally polarized flux from a circumstellar disk appears as a consistently positive signal in the $Q_\phi$ image, whereas the $U_\phi$ image remains free of disk
signal and provides a convenient estimate of the residual noise in the $Q_\phi$ image  \citep{schmid2006}.  
Because the DPI mode of IRDIS was still in commissioning at the time of publication, the absolute level of polarization could not be accurately retrieved, and we are only able to report polarized intensity images with arbitrary surface brightness levels.

\section{Polarized intensity images}
\label{sec:images}
Figure~\ref{fig:Qphi} shows the $Q_\phi$ images at both epochs and the radial mapping of the March data, obtained after deprojection \citep[with $i=21^\circ$, $PA=65^\circ$;][]{isella2010}. Each pixel was scaled with $r^2$ to compensate for the $r^{-2}$ dependency of the stellar illumination. The dark central region corresponds to the area masked by the coronagraph. The residual spiders from the coronagraph are visible in the December 2014 image (Fig.~\ref{fig:Qphi}, left). Scattered light is detected from $\sim$0.093\arcsec{} to $\sim$0.53\arcsec{} ($\sim$26 to $\sim$148\,au) in polarized intensity, and we did not resolve any depleted cavity.  Six distinct non-axisymmetric features are detected, the innermost ones being at previously inaccessible radii.  The features are indicated by arrows in Fig.~\ref{fig:Qphi}: (1)~a spiral arm located in the south and east (hereafter, the SE spiral) at deprojected distances $r\sim$0.26\arcsec{}--0.53\arcsec{}; (2)~an arc in the west at $r\sim$0.2\arcsec{}--0.3\arcsec{}; (3)~a spiral arm  in the northwest (hereafter, the NW spiral) at $r\sim$0.22\arcsec{}--0.34\arcsec{}; (4)~an arc located north of the NW spiral at $r\sim$0.32\arcsec{}--0.53\arcsec{};  (5)~substructures inside the spirals at $r\sim$0.1\arcsec{}--0.2\arcsec{} and (6)~a substructure branching off the SE spiral at $r\sim$0.21\arcsec{}--0.27\arcsec{}. We note that although there is some clear scattered light signal very close to the coronagraph, the exact morphology of feature (5) may be affected by the efficiency of the AO correction and the coronagraph centering. Appendix~\ref{sec:snr} provides the $U_\phi$ images (Fig.~\ref{fig:Uphi}) and SNR maps (Fig.~\ref{fig:snr}). The features are detected above the disk background at more than 3-$\sigma$ and do not appear in the $U_\phi$ images. This suggests that they are real and not diffraction residuals that would affect the $Q_\phi$  and $U_\phi$ images alike. 
Although one could naturally see a single spiral arm in the west, extending from $PA\sim180^\circ$ to $350^\circ$, the radial mapping (Fig.~\ref{fig:Qphi}, right)  shows that it consists of different features that have a very distinct dependence on the $PA$, with a sharp transition at $PA\sim240^\circ$. In the December 2014 image, feature (2) hosts an area that is $\sim$1.5 times brighter than the NW spiral, at $PA\sim240^\circ-260^\circ$. In the March dataset, the area at $PA\sim260^\circ$ in contrast seems fainter than the background spiral.  Since it is very close to the residual spiders, it is difficult to conclude whether it is an instrumental effect, or if it has a physical origin which, considering the timescale of the variations, would point to a variability in the inner disk.
While the detected structures agree with the HiCIAO $H$-band image \citep{grady2013}, we did not detect a significant difference in the extent of the scattered light signal between the west and the east sides. 


\section{Spiral feature modeling} 
\label{sec:spi}
Planets embedded in circumstellar disks are known to launch spiral waves at Lindblad resonances both inside and outside of their orbit \citep[e.g.][]{ogilvie2002}. The shape of the spiral wake is determined by the disk rotation profile and by the sound speed (i.e., temperature) distribution in the disk.  In this section, we attempt to fit the shape of the two regions where there is a significant departure from circular symmetry, features (1) and (3), with a model based on the spiral density wave theory. As the  spiral features are very similar in both datasets, and  since the observations in March were obtained under better conditions, we only fit the corresponding dataset.  
We assume that the observed scattered light traces small dust grains that are well coupled to the gas, and thus, that it indirectly traces the gas. Following the prescription of \citet{rafikov2002} and \citet{muto2012}, the spiral wake in polar coordinates ($r, \Phi$) follows


 \begin{eqnarray}
\nonumber \Phi(r) &&= \Phi_{c} +  \frac{{\rm sgn}(r-r_{c})}{h_{c}} \left(\frac{r}{r_{c}} \right)^{1+\beta} \left(\frac{1}{1+\beta} - \frac{1}{1-\alpha+\beta}\left(\frac{r}{r_{c}}\right)^{-\alpha}\right) \\
 &&- \frac{{\rm sgn}(r-r_{c})}{h_{c}} \left( \frac{1}{1+\beta} - \frac{1}{1-\alpha+\beta}\right).
\label{eq:sp_wake_muto}
\end{eqnarray}

\noindent where $\alpha$ and $\beta$ are the exponents of the disk rotation and sound speed profiles, respectively: $\Omega\propto r^{-\alpha}$, $c_s \propto r^{-\beta}$. $h_c = c_s(r_c)/r_c\Omega(r_c)$ is the ratio of the pressure scale height $H$ to the radius (also called disk aspect ratio) at the location of the planet ($r_c, \Phi_c$). The pitch angle of the spirals depends on the disk temperature (hence on the aspect ratio) and the distance from the launching planet. The flaring index,  $\alpha - \beta -1$, determines the variation of $H/r$ with radius. 
Equation~\ref{eq:sp_wake_muto} is valid in the linear or weakly nonlinear regimes, in which a single spiral wave is launched by an embedded planet, and approximates well the shape of the density wave given by the WKB theory \citep{rafikov2002}. 

We attempted to fit both spirals simultaneously assuming that they are launched by two planets in circular orbit at different radii in the disk.  We assumed that the disk is in Keplerian rotation and fixed $\alpha$=3/2. 
Varying $\beta$ has little influence on the fit, and we fixed $\beta$=0.45, following \citet{andrews2011}. 
We considered that $h_c$ is a global disk parameter, meaning that the values of aspect ratios at the locations of the planets should be consistent.  This leads to five free parameters in total. We restricted our models to planets located inside the submillimeter cavity, whose extent is $\sim$0.36\arcsec{} within large uncertainties  \citep{isella2010, andrews2011} and considered disk aspect ratios of at most 0.20 to be consistent with the modeling of the spectral energy distribution (SED) \citep{andrews2011}. The parameters are thus varied as $0.03\leq h_c \leq 0.20$, $0.01\arcsec{}\leq r_{c,\rm{NW/SE}} \leq 0.36\arcsec{}$, $0\leq \Phi_{c,\rm{NW/SE}}\leq 360 ^\circ$, in 20, 50, and 50 linearly spaced values, respectively. 



\begin{figure}[t]
   \centering
    \includegraphics[width=0.5\textwidth]{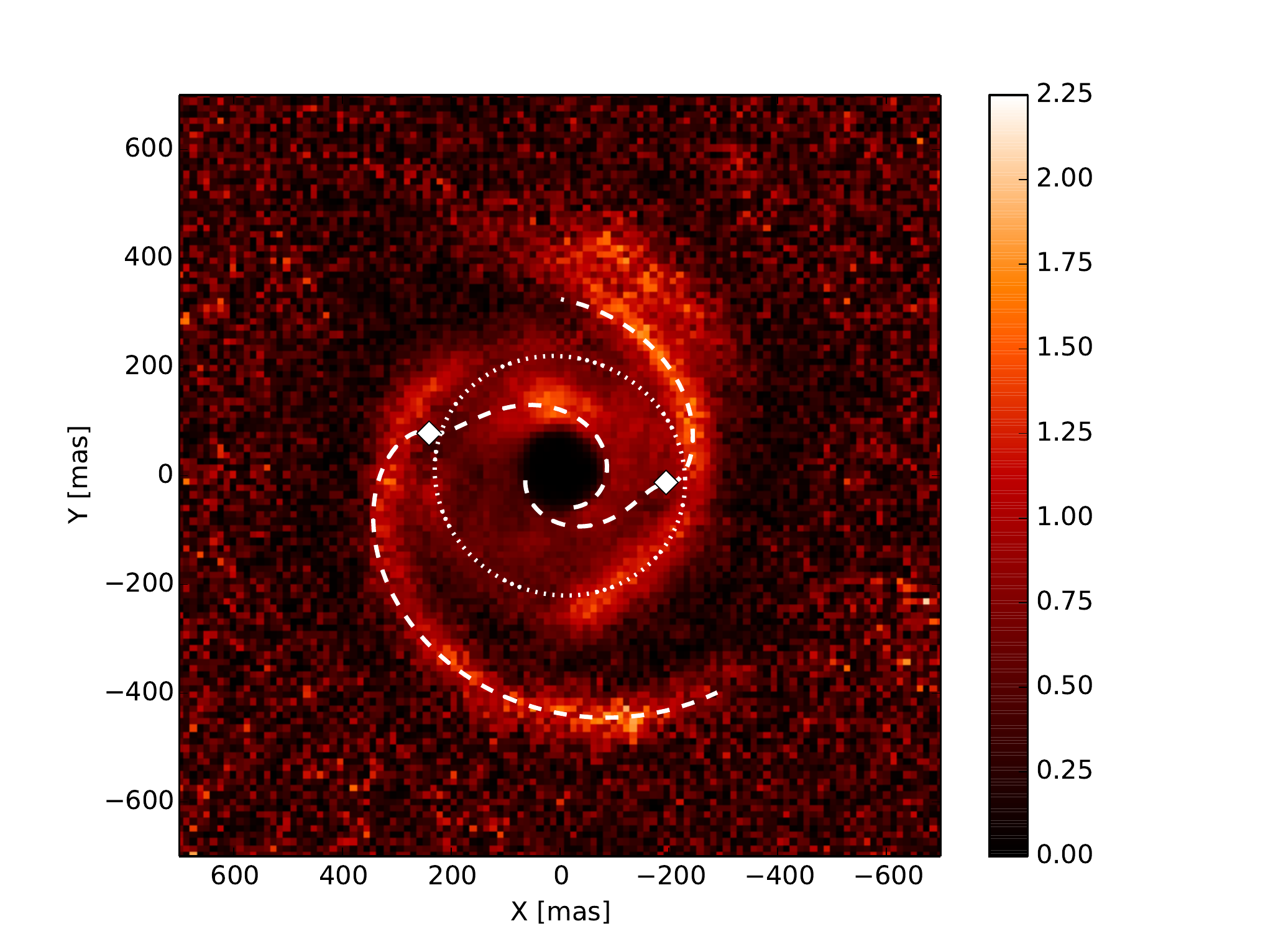}	
   \caption{Best-fit model in dashed lines. The locations of the planets are indicated by  white diamonds. The dotted ellipse is a projected circle with radius 0.25\arcsec{}. The color scale is arbitrary.}   
   \label{fig:bestfitboth}
\end{figure}

 
We  deprojected the image and fit the locations of the surface brightness maxima  along a set of azimuth angles ($PA\sim$70$^\circ$--200$^\circ$ (80 linearly spaced values) and 280$^\circ$--340$^\circ$ (50 values), for the SE and NW spirals, respectively). We minimized a $\chi^{2}$ function assuming that the uncertainty on each location is the FWHM of the point spread function (PSF). The best-fit parameters are for the SE arm: ($h_c$, $r_c$, $\Phi_c$) = (0.20, 0.253\arcsec{}, 72$^\circ$), and for the NW arm:  
 ($h_c$, $r_c$, $\Phi_c$) = (0.20, 0.196\arcsec{}, 266$^\circ$). 
This corresponds to planets located at $\sim$55 and 71\,au for the NW and SE arms, respectively. The spiral shapes with these parameters are plotted in the polarized intensity image in Fig.~\ref{fig:bestfitboth}.  As noted by \citet{muto2012}, the degeneracy of the model is significant. Figure~\ref{fig:probboth} shows the 2D probability distributions for each pair of free parameters. The Bayesian probabilities were derived as $\rm{exp}(-\chi^2/2)$ and normalized such that their integral over the whole parameter domain is unity. We obtained a lower limit of the disk aspect ratio $h_c$=0.2. A high value of $h_c$ is needed to account for the large pitch angle, and while higher values may be favored by the spiral fit, they would produce too much infrared flux to account for the SED. In addition, we note that $h_c$ and $r_c$ are not independent. The lower $h_c$, the closer to the spiral the planet must be, as the spiral pitch angle is only large very close to the planet.

Although we restricted the planets to be inside the submillimeter cavity, we note that the only way to fit the northern tip of the SE spiral (up to $PA\sim$45$^\circ$) is to consider a planet outside of the  submillimeter  continuum emission as in  \citet{grady2013}. 
However, extending the range in $r_c$ to 2\arcsec{} strongly increases the degeneracy of the model. We do not favor this solution because companions are likely to be located inside the submillimeter cavity since other clearing mechanisms, such as photo evaporation, can be ruled out by near-IR interferometric observations of the inner disk \citep{isella2008}.

\section{Discussion and conclusions}
\label{sec:dis}
We presented $Y$-band polarized intensity images of the scattered light  from the protoplanetary disk around the Herbig Ae star MWC~758 and studied the morphology of non-axisymmetric features. The image shows six distinct features, including arc-like features and two spiral arms with a large winding angle. Fitting the shape of the spirals, with a model based on the spiral density wave theory, showed that only a warm disk with a high aspect ratio ($\sim$0.2) at the locations of the planets can lead to such large pitch angle spirals \citep{rafikov2002}. Assuming that the disk is in hydrostatic equilibrium, this corresponds to midplane temperatures of $\sim$315\,K and $\sim$250\,K at 55 and 71\,au, respectively, slightly higher than the predictions from SED and submillimeter continuum modeling \citep[$\sim$200\,K at these locations;][]{andrews2011}, but significantly different from the temperature measured with CO \citep[$\sim$53~K at 50 AU;][]{isella2010}. Such a high disk aspect ratio also reduces the contrast and the observability of the spirals in scattered light \citep{juhasz2014}.  This suggests that the vertically isothermal assumption implied in the spiral model may not hold \citep[e.g.][]{richert15}.  Alternatively, a planet on an eccentric orbit, an inclined planet or a planet massive enough to exert disk eccentricity may affect the shape of the spirals and induce a different pitch angle than in the case of a circular orbit.

The two spiral arms were interpreted in the context of planet-disk interaction, but other mechanisms, such as self-gravity, can trigger spiral features with lower winding numbers \citep{ricelodato2006}, while keeping enough contrast to detect them in scattered light (Pohl et al, in prep.) and at longer wavelengths \citep{dipierro2014}. The disk mass has been estimated to be $\sim$0.01~M$_{\odot}$ from submillimeter observations \citep{andrews2011}, which is probably too low to trigger gravitational instabilities, although these  estimates are very uncertain and strongly depend on the assumed dust opacities and gas-to-dust ratio. If a significant part of the solid mass is in planets or particles larger than 1\,cm, the gas mass can be much higher than currently estimated.

Assuming that the observed spirals are trailing, the SE spiral arm is located on the far side of the disk and interestingly, does not appear darker than the near side \cite[unlike SAO206462,][]{garufi2013}. This may indicate an isotropic scattering phase function of the dust grains, which would imply submicron sized grains, or a high polarization efficiency. We note that the detected scattered light  is quite symmetric and the western side of the image does not show the extended emission seen in the HiCIAO image \citep{grady2013}, which was interpreted as being due to an asymmetric irradiation of the disk. Considering that the observations were obtained only three years apart, it is unlikely that this would result from variable shadowing and should be investigated further. 

We found no indication of a fully depleted cavity in micron-size dust grains beyond the coronagraph radius \citep[consistent with $^{12}$CO peaking at the star position;][]{isella2010} while a cavity in millimeter grains has been marginally resolved at a much larger radius \citep{isella2010,andrews2011}. Such spatial segregation of small and large grains can be a natural outcome of particle trapping at the edge of a cavity carved by a planet \citep{pinilla2012,pinilla2015}. The requirement of a continuous replenishment of small particles through the cavity translates into a maximum planet mass, above which any companion would filter all dust particles \citep{rice2006,zhu2012}. With typical disk viscosities, a companion with a planet-to-star mass ratio above 10$^{-2}$ (i.e., a 5.5~$M_{\rm{Jup}}$ planet around MWC~758) would filter all dust grains \citep{pinilla2012}.

Features (2) and (3) have a very distinct dependence on the $PA$, with a sharp transition at $PA\sim$240$^\circ$, therefore no spiral model can fit the full NW structure (features (2) and (3) together).  A projected circle of 0.25\arcsec{} radius is shown in Fig.~\ref{fig:bestfitboth} and can only partly account for the region between $PA\sim$190-270$^\circ$.  The circular region between features (2) and (3) may trace the edge of a cavity, although high-resolution submillimeter observations are required to determine whether this is the case. Alternatively, a slightly eccentric ellipse (e$\sim$0.1) oriented with $PA$=-30$^\circ$, that would trace an eccentric gap, can fit part of feature (3) in addition to feature (2). 
In addition to the two spiral arms, four more structures are detected. Feature (4) is located beyond the NW spiral and indicates that the NW arm does not shadow the outer disk in this direction. Its small angular extent is puzzling, and its nature is unclear.  Features (5) and (6) are located closer in than the two arms and the submillimeter cavity, at previously inaccessible radii close to the star. Feature (5) may trace the  optically thick innermost disk \citep{isella2008}, or alternatively, the inner arm of the SE spiral.

Considering that the best-fit aspect ratio from the spiral models is difficult to reconcile with the expected temperatures in this disk, and that on the other hand, the disk mass seems to be too low to account for gravitational instability, additional observations coupled with advanced modeling are needed to understand which processes these non-axisymmetric features may trace. In particular, complementary observations on similar spatial scales with ALMA, which will trace different optical depths in the disk, is  very promising. Future DPI observations with SPHERE and its visible-light instrument ZIMPOL \citep{zimpol2010} are also expected to place new constraints on the innermost structures. 


\begin{acknowledgements}
We acknowledge the SVT team at ESO HQ for their help during the preparation of the OBs and the VLT team for conducting the observations. We thank C.P.~Dullemond, G.~Lesur, M.~Min, and M.~Tauras for fruitful discussions, and the referee for providing useful comments. M.B. acknowledges financial support from "Programme National de Physique Stellaire" (PNPS) of CNRS/INSU, France. C.G. was supported under the NASA Origins of Solar Systems program on NNG13PB64P. T. B. acknowledges support from NASA Origins of Solar Systems grant NNX12AJ04G.
\end{acknowledgements}

\bibliographystyle{aa}
\bibliography{mwc758}

\appendix
\section{$U_{\phi}$ image and SNR map}
\label{sec:snr}

\begin{figure*}[t]
   \centering
   		\includegraphics[width=1.0\textwidth]{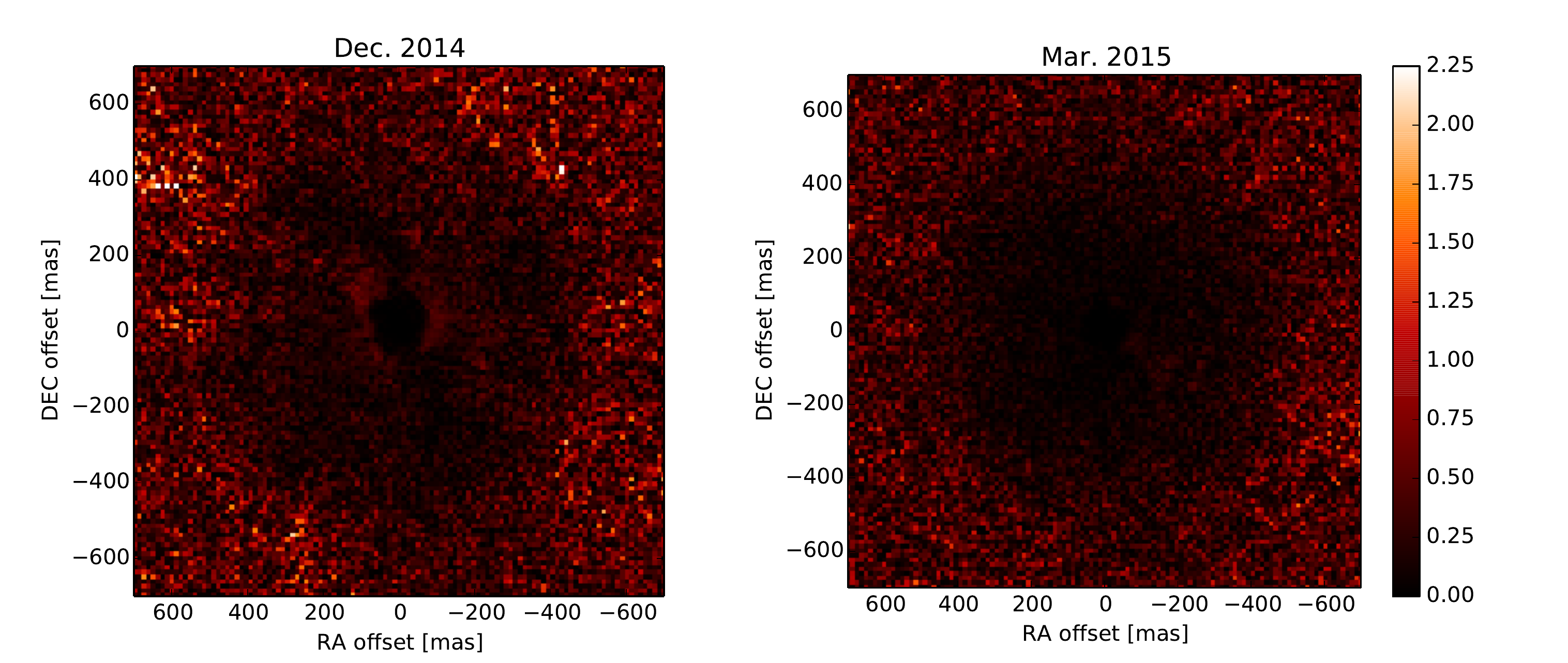}  
	\caption{$U_{\phi}$ images (left, December 2014; right, March 2015). East is toward the left. Each pixel has been scaled with the square of its distance from the star, $r^{2}$, to compensate for the $r^{-2}$ dependency of the stellar illumination. The color scale is arbitrary.}
		\label{fig:Uphi}
\end{figure*}

Figure~\ref{fig:Uphi} presents the $U_\phi$ images. Since the scattered light from a circumstellar disk is expected to be linearly polarized in the azimuthal direction under the assumption of single scattering, $U_\phi$ contains no signal, but only noise of the same magnitude as the noise in the $Q_\phi$ image. For multiple scattering, the assumption of polarization only in azimuthal direction still holds approximately, especially for close to face-on disks, and the deviations from azimuthal polarizations are expected to be well below the noise level \citep{avenhaus2014}.

\begin{figure*}[t]
   \centering
   \includegraphics[width=1.0\textwidth]{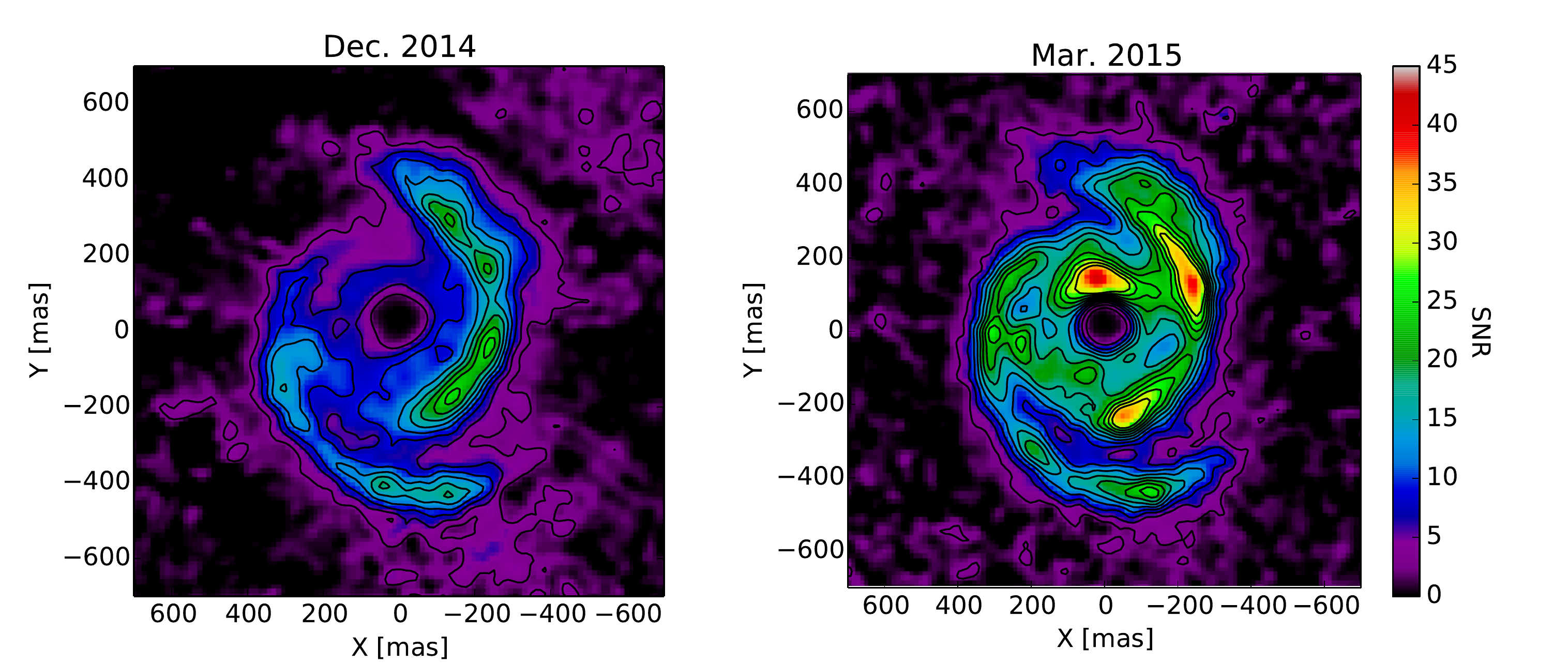}	
   \caption{SNR maps with contours every 3~$\sigma$ (left, December 2014; right, March 2015)}. 
   \label{fig:snr}
\end{figure*}

Figure~\ref{fig:snr} provides qualitative SNR maps. They were obtained by first smoothing both $Q_\phi$ and $U_\phi$ with a Gaussian kernel of 37 mas (close to the instrument PSF) and then dividing the $Q_\phi$ by the standard deviation in the $U_\phi$ images. The standard deviation in the $U_\phi$ image was calculated over a Gaussian-weighted area centered on the respective image point with a  FWHM of 260 mas.  We note that these do not strictly provide signal-to-noise ratio estimates. Indeed, the noise reported in these maps depends on the area considered for the calculation of the variance from the $U_\phi$ images. Moreover, the first step smoothes out the readout noise and other small-scale (smaller than the smoothing kernel) noise (but not large-scale systematic noise). These maps are still very useful, as a feature detected in the $Q_\phi$ image and appearing at a location of the map that has a high SNR value, should be considered as real. An example is feature 4, which lies in an area of high SNR and can thus be considered real, even though it does not seem apparent from the SNR maps alone.

\section{Probability plot}
\label{app:proba}
\begin{figure*}[t]
   \centering
   \includegraphics[width=1.0\textwidth]{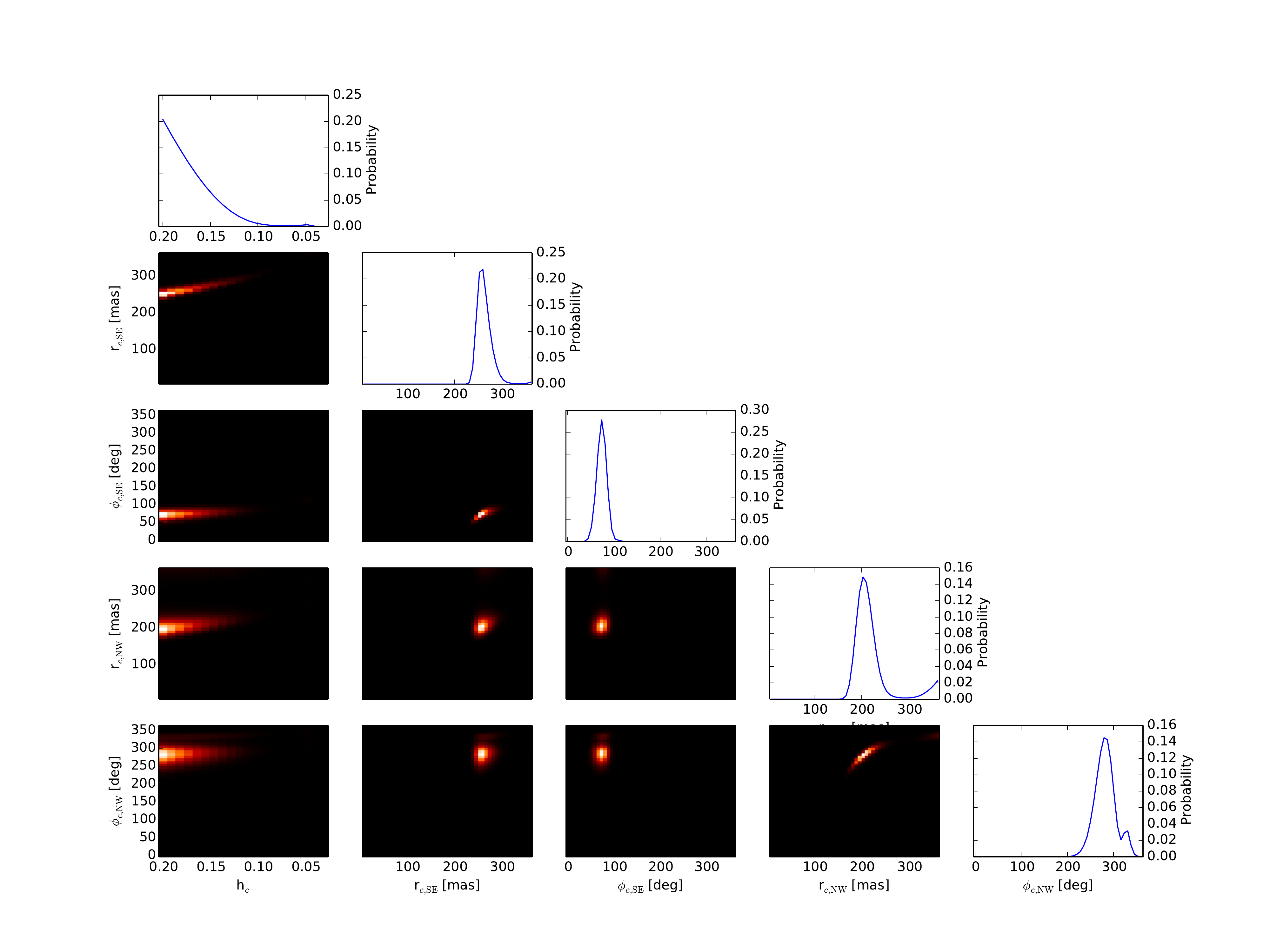}	
   \caption{Bayesian probabilities for the simultaneous fit of the two spirals. The top panel of each column provides the integrated 1D probability for each parameter. The lower panels in each column are the 2D probability maps for each pair of parameters.}
   \label{fig:probboth}
\end{figure*}

Figure~\ref{fig:probboth} shows the 2D probability distributions for each pair of free parameters. 

\end{document}